\input ./style/arxiv-general.cfg
\documentclass[aoas,MSNbibl,nameyear,seceqn,dvips]{arximspdf}
\makeatletter
   \@ifpackageloaded{graphicx}{}{\usepackage{graphicx}}
\makeatother
\usepackage{dcolumn}
\usepackage{url,breakurl}

\doi{10.1214/15-AOAS872}
\volume{9}
\issue{4}
\pubyear{2015}
\firstpage{1889}
\lastpage{1905}
\docsubty{FLA}

\makeatletter
\newcolumntype{d}[1]{D{.}{.}{#1}}
\newcommand{\eqref}[1]{(\ref{#1})}
\newcommand{\by}{\mathbf{y}}
\newcommand{\btheta}{\bolds{\theta}}
\newcommand{\bz}{\mathbf{z}}
\newcommand{\bQ}{\mathbf{Q}}
\newcommand{\bmzero}{\mathbf{0}}

\makeatother

\begin{document}
\begin{frontmatter}

\title{Space--time smoothing of complex survey data: Small area
estimation for child mortality\thanksref{T1}}
\runtitle{SAE of child mortality}
\thankstext{T1}{Supported in part by USAID through the MEASURE DHS
project and by a Shanahan Endowment Fellowship and a Eunice Kennedy
Shriver National Institute of
Child Health and Human Development (NICHD) training grant, T32
HD007543,
to the Center for Studies in Demography \& Ecology at the University of
Washington and a 2R01CA095994-05A1 from the National Institutes of
Health.}

\begin{aug}
\author[A]{\fnms{Laina D.} \snm{Mercer}\thanksref{m1}\ead[label=e1]{mercel@uw.edu}},
\author[B]{\fnms{Jon} \snm{Wakefield}\corref{}\thanksref{m1}\ead[label=e2]{jonno@uw.edu}},
\author[C]{\fnms{Athena} \snm{Pantazis}\thanksref{m1}\ead[label=e3]{apantazi@u.washington.edu}},
\author[D]{\fnms{Angelina~M.} \snm{Lutambi}\thanksref{m2}\ead[label=e4]{alutambi@ihi.or.tz}},
\author[D]{\fnms{Honorati} \snm{Masanja}\thanksref{m2}\ead[label=e5]{hmasanja@ihi.or.tz}}
\and\\
\author[C]{\fnms{Samuel} \snm{Clark}\thanksref{m1,m3,m4,m5,m6,T2}\ead[label=e6]{samclark@u.washington.edu}}
\runauthor{L.~D. Mercer et al.}
\affiliation{University of Washington\thanksmark{m1},
Ifakara Health Institute\thanksmark{m2},
University of Colorado\thanksmark{m3},
University of the Witwatersrand\thanksmark{m4},
INDEPTH Network\thanksmark{m5} and
ALPHA Network\thanksmark{m6}}
\thankstext{T2}{Supported in part by Grant K01 HD057246 from
the NICHD.}
\address[A]{L.~D. Mercer\\
Department of Statistics\\
University of Washington\\
Seattle, Washington 98195\\
USA\\
\printead{e1}}
\address[B]{J. Wakefield\\
Department of Statistics\\
Department of Biostatistics\\
University of Washington\\
Seattle, Washington 98195\\
USA\\
\printead{e2}}
\address[C]{A. Pantazis\\
S. Clark\\
Department of Sociology\\
University of Washington\\
Seattle, Washington 98195\\
USA\\
\printead{e3}\\
\phantom{E-mail:\ }\printead*{e6}}
\address[D]{A.~M. Lutambi\\
H. Masanja\\
Ifakara Health Institute\hspace*{11pt}\\
Dar es Salaam\\
Tanzania\\
\printead{e4}\\
\phantom{E-mail:\ }\printead*{e5}}
\end{aug}

%
\received{\smonth{11} \syear{2014}}
%
\revised{\smonth{9} \syear{2015}}

%
\begin{abstract}
Many people living in low- and middle-income countries are not covered
by civil registration and vital statistics systems. Consequently, a
wide variety of other types of data, including many household sample
surveys, are used to estimate health and population indicators. In
this paper we combine data from sample surveys and demographic
surveillance systems to produce small area estimates of child mortality
through time. Small area estimates are necessary to understand
geographical heterogeneity in health indicators when full-coverage
vital statistics are not available. For this endeavor spatio-temporal
smoothing is beneficial to alleviate
problems of data
sparsity. The use of conventional hierarchical models requires careful
thought since the survey weights may need to be considered to
alleviate bias due to nonrandom sampling and nonresponse. The
application that motivated this work is an estimation of child mortality
rates in five-year time intervals in regions of Tanzania. Data come
from Demographic and Health Surveys conducted over the period
1991--2010 and two demographic surveillance system sites. We derive a
variance estimator of under five years child mortality that accounts
for the complex survey weighting. For our application, the
hierarchical models we consider include random effects for area, time
and survey and we compare models using a variety of measures including
the conditional predictive
ordinate (CPO). The method we propose is implemented via the fast and
accurate integrated nested Laplace approximation (INLA).
\end{abstract}

%
\begin{keyword}
\kwd{Bayesian smoothing}
\kwd{infant mortality}
\kwd{small area estimation}
\kwd{survey sampling}
\end{keyword}
\end{frontmatter}

\section{Introduction}\label{sec1}

Over the past fifteen years the United Nations' (UN) Millennium
Development Goals (MDGs) [\citet{mdgsWeb}] have focused the world's
attention on improving key indicators of development, health and
wellbeing. The requirement to monitor progress toward the MDGs has
revealed a stunning absence of data with which to measure and monitor
key indicators related to the MDGs in much of the developing world,
and this has led to great interest in improving both the data and our
ability to use it. In 2015 the UN and its partners are
taking stock of experience with the MDGs and coordinating the
establishment of a new set of global goals [\citet{post2015Web}]---the
Sustainable Development Goals (SDGs) [\citet{sdgsWeb}]. Even before the
SDGs are finalized, the UN Secretary General has called for a
\textit{Data Revolution for Sustainable Development} and appointed a
high-level advisory group to define what it should be
[\citet{dataRevWeb}]. The aim is clear: to rapidly improve the
coverage, quality, availability and timeliness of the data used to
measure and monitor progress toward the SDGs. Simultaneously, there is
sustained, strong interest in improving civil registration, vital
statistics (CRVS) and the functioning of statistical offices across
the developing world [\citet{crvsScaleUpWeb, paris21Web}]. The key
challenges are improving coverage [\citet{crvsCoverageWeb}] and
timeliness of reporting.

In this context of far-reaching interest in improving data and methods
available to monitor indicators of the SDGs and improve CRVS, in this
paper we
develop a general approach that combines data from different sources
and provides temporal, subnational-specific estimates with uncertainty that
accounts for the different designs of the data collection schemes. We
demonstrate the method by calculating spatio-temporal estimates of
child mortality in Tanzania using data from
multiple Demographic and Health Surveys (DHS) [\citet{dhs}] and two
health and
demographic surveillance system (HDSS) sites [\citet{dss}].

Reducing child mortality is MDG 4 [\citet{mdg4}], and over the past
fifteen years a great deal of effort and resources have been spent in
order to
meet MDG 4 targets at the national level in many developing nations.
This has driven work to develop better methods to estimate trends in
child mortality at the national level, and two groups have produced
globally comparable trends in child mortality for all nations. The
United Nations Inter-agency Group for Child Mortality Estimation (UN
IGME) recently developed a Bayesian B-spline Bias-reduction (B3) method
[\citet{alkema2014child,alkemanew:aoas:2014}], and the Institute for
Health Metrics and Evaluation (IHME) uses a
Gaussian process regression [\citet{wang:etal:2014}].
Both of these methods produce national
estimates through time with measures of uncertainty. None are designed to
reveal variation in child mortality within countries. A recent paper by
\citet{dwyer:etal:2014} compared many Bayesian space--time smoothing
models to produced subnational
estimates of U5MR for Zambia. The major methodological limitation of
this approach is that it does not incorporate area-specific
sampling variability at the first stage of analysis, which we show can
be quite variable for small areas. 

In this paper we combine data from multiple surveys
with different sampling designs, and construct subnational
estimates through time with uncertainty that reflects the various
data collection schemes. Data come from traditional cluster
sample surveys (DHS) and two HDSS sites. HDSS sites intensively monitor
everyone within a given area, typically to monitor the effects of
health intervention trials of various types. Estimates of child
mortality from both sources of data are useful but potentially flawed
in different ways. National cluster sample surveys are generally not
able to produce useful subnational estimates, and HDSS sites are not
designed to be nationally representative, and are also thought to fall
prey to the Hawthorne effect by which the communities of these sites
have improved health outcomes because they are under observation and,
more concretely, because of the trials
being conducted.

We construct subnational estimates of Tanzanian child mortality
through time with uncertainty intervals. This problem is challenging
because in addition to requiring smoothing over space and time, we
must also account for the survey design. When sampling is not simple
and random and the design variables (upon which sampling was based)
are not available, the complex sampling design is accounted for by
constructing design weights. Inference is then carried out using
design-based inference, for example, using Horvitz--Thompson estimators
[\citet{horvitz:thompson:52}].
In contrast, a conventional space--time random effects framework, for
example, \citet{knorrheld:00},
is model based, and requires an explicit likelihood to be specified. In
this paper, we marry these two approaches by constructing a working
likelihood based on the asymptotic distribution of a design-based
estimator and then smooth using a space--time--survey hierarchical prior.

The organization of this paper is as follows. In Section~\ref{sec:data}
we describe the two data sources upon which estimation will be based.
In Section~\ref{sec:dtsa} the calculation of child mortality estimates
with an appropriate standard error is described using discrete time
survival models. Hierarchical Bayesian space--time models are introduced
in Section~\ref{sec:models}.
The results of our modeling efforts of under five mortality rates
(U5MR) within Tanzania from 1980--2010 are given in Section~\ref{sec:modresults} and discussed in Section~\ref{sec:discuss}.

\section{Data sources}
\label{sec:data}

We focus on child mortality using data from five Tanzanian Demographic
and Health Surveys (TDHS): one Tanzania HIV and Malaria Indicator
survey (THMIS), and two health and demographic
surveillance system (HDSS) sites in Tanzania, Ifakara and Rufiji. Over
the period 1980--2010 estimates of child mortality from the two types
of data sources (surveys, surveillance sites) are generally similar
but, as described above, different
in useful ways. The HDSS estimates are accurate (low bias) and precise
(small variance)
measurements for comparatively small, geographically-defined
populations, and
the household survey estimates are less accurate and much less precise
but representative of large
populations.

\subsection{Health and demographic surveillance system}

The Ifakara Health Institute (IHI), Tanzania runs a number of health
and population research projects including two HDSS sites---Ifakara
and Rufiji. We collaborated with IHI to estimate child mortality using
data from the Ifakara and Rufiji HDSS sites.

The HDSS data are generated through repeated household visits. For the
data we use, each household was visited three times per year at regular
intervals. During each visit a ``household roster'' was updated and all
new vital and migration events for all members of the household were
recorded. In addition, potentially many other questions were asked as
part of both routine and ``add-on'' studies. For our purposes we
require only the basic core HDSS data that include information on dates
of birth, death and migration---the information necessary to
accurately identify observed person time, categorize that time by
calendar period and age, and identify the outcome of interest, death.
The Ifakara and Rufigi HDSS sites contribute data to the Morogoro and
Pwani regions of Tanzania, respectively.

\subsection{Household surveys}

Full TDHS surveys that collected data necessary for child mortality
estimates were conducted in Tanzania in 2010, 2004--2005, 1999, 1996 and
1991--1992, in addition to the THMIS that included child mortality which
was conducted in 2007--2008. The 2010 TDHS, 2007--2008 THMIS and 2004--2005
TDHS surveys used 2-stage cluster samples. First, enumeration areas
were sampled from the 2002 Tanzania census and, second, a systematic
sampling of households within each enumeration area was carried out.
The 1999 TDHS, 1996 TDHS and 1991--1992 TDHS used a 3-stage cluster
design, first selecting wards and branches using the 1988 Tanzania
Census as a sampling frame, second using probability proportional to
size sampling to select enumeration areas from each selected ward or
branch, and third selecting households from a new list of all
households in each selected enumeration area. The same first and second
stage units were used for all three of the surveys. For all surveys
stratification by urban/rural and region was done at the first stage,
with oversampling of Dar es Salaam and other urban areas. The surveys
were designed to be nationally representative and to be able to provide
estimates of contraceptive prevalence at the regional level. All six
household surveys contributed observations to the 21 mainland regions
of Tanzania.

All women age 15 to 49 who slept in the household the night before
were interviewed in each selected household and response rates were
high (above 95\% for households in all surveys). TDHS provides sampling
(design) weights, assigned to each individual in the data set. Limited
information is provided for each survey concerning the calculation of
survey weights, but the general explanation indicates that raw survey
weights are the inverse of the product of the 2--3 probabilities of
selection from each stage. These raw weights were then adjusted to
reflect household response and individual response rates. The 1991--1992
Tanzania DHS final report [\citet{TanzaniaDHS:91}] states that ``final
individual weights were calculated by normalizing them for each area so
that the total number of weighted cases equals the total number of
unweighted cases,'' but this normalization is not discussed in later
reports [\citeauthor{TanzaniaDHS:04} (\citeyear{TanzaniaDHS:96,TanzaniaDHS:99,TanzaniaDHS:04,TanzaniaDHS:10})] or the
DHS statistics manual [\citet{DHSguide06}].
For the purposes of our analysis of child mortality, children
identified by the women who were interviewed contributed exposure time
and deaths. The data were organized into child-months from birth to
either death or date of the mother's interview.

\section{Calculating child mortality with discrete time survival models}
\label{sec:dtsa}


We modeled child mortality using discrete time survival analysis (DTSA)
[\citet{allison1984event,jenkins1995easy}]. Our main aim is to examine
the change in risk as a function of age and historical period. DTSA
allows us to easily estimate the predicted probabilities which can be
used directly in traditional mortality analysis methods such as life
tables, in our case to calculate U5MR. We wish to estimate the U5MR and
define $_{n}q_x=\Pr(\mbox{dying before } x+n |  \mbox{lived until
}x)$ and the discrete hazards model splits the $[0,5)$ period into $J$
intervals $[x_1,x_2),[x_2,x_3),\ldots,[x_{J},x_{J+1})$, where
$x_{j+1}=x_j+n_j$ so that $n_j$ is the length of the interval beginning
at $x_j$, $j=1,\ldots,J$.
Then U5MR is calculated as
%
\begin{equation}
\label{eq:dis1} _5q_0 = 1- \prod
_{j=1}^J (1- {_{n_j}q_{x_{j}}} ).
\end{equation}
For our purposes, $_5q_0$ is calculated by dividing the first 60 months
into six intervals ($J=6$),
$[0,1),[1,12),[12,24),[24,36),[36,48),[48,60)$ with
$(x_1,\ldots,\break x_6)=(0,1,12,24,36,48)$ and
$(n_1,\ldots,n_6)=(1,11,12,12,12,12)$. Data were organized as child-months
where each child was at risk during each month observed from birth up
to and including the month of their death. The observed data consist
of, for each birth, a binary sequence up to length 60 with 0/1
corresponding to survival/death. For example, a child that died in
their fourth month would contribute one child-month to the first age
category and three to the second age category. The first three
child-months would be assigned a 0 outcome and the final month would be
assigned a 1.

We use logistic regression to estimate the monthly probability of dying
conditional on the state of the child at the beginning of the month.
The monthly probability of death for each interval, $_1q_x$, is the
probability of dying in $[x,x+1)$ for $x\in[x_j,x_j+n_j)$ and
can be estimated using a logistic generalized linear model (GLM) with
$J$ factors for age intervals, $\operatorname{logit} ( _1q_x  )=
\beta
_j $ for $ x \in[x_j,x_j+n_j)$. A~more detailed discussion of the DTSA
method can be found in \citet{clark2013young}.

In the complex survey context that is relevant for the Tanzanian
household surveys, an important consideration is that the design
weights must be acknowledged. This is achieved by solving a (design)
weighted score statistic [\citet{binder:83}], resulting in estimates of
the finite population parameter $\mathbf{B}=[B_1,\ldots,B_J]$; see
details in the supplementary material [\citet{mercer:etal:2015supp}].
Once $\widehat{B}_j$ are estimated, we can calculate $\widehat
{_1q_x}=\exp(\widehat{B}_j)/[1+\exp(\widehat{B}_j)]$ for $x\in
[x_j,x_j+n_j)$. The complement of surviving each month of the interval
$[x_j,x_j+n_j)$ is used to calculate ${_{n_j}\widehat{q}_{x_{j}}} =
1- (1-\widehat{_1q_x}  )^{n_j}$, which may be substituted into
(\ref{eq:dis1}) to give $_5\widehat{q}_0$ (for additional details see
the supplementary material).\vadjust{\goodbreak}

In Section~\ref{sec:models} we will construct, for a generic U5MR, a
working likelihood based on the asymptotic distribution
\[
y = \operatorname{logit} ( _5\widehat{q}_0 ) \sim\mathrm{N}(\eta
,\widehat{V}_{\mathrm{{DES}}}),
\]
where $\eta=\log[ _5q_0/(1-_5q_0)]$ and $\widehat{V}_{\mathrm{
{DES}}}$ is the estimated asymptotic (design-based) variance estimate of $\operatorname
{logit} ( _5\widehat{q}_0 ) $, which is obtained via the delta
method; the supplementary material contain details of this calculation
and a simulation study which investigates the asymptotic properties of
the variance estimate compared with a jackknife variance estimate
[\citet{lohr:2009}, Chapter~9] that is often used in the in the context
of child mortality estimates [\citet{pedersenandliu:2012}].

Simulation results were much as one would expect from clustered
sampling; coverage improves when there are more clusters and within a
given number of clusters there is little gain in precision when
increasing the sample size. Generally the performance of the delta
method and jackknife intervals is very similar. We prefer the delta
method, as it is generally applicable (i.e., to a variety of designs)
and has a far smaller computational burden. We conclude that the
asymptotic normal sampling distribution and the delta method variance
result in sufficiently accurate confidence interval coverage for the
cluster and sample sizes considered in our application. Consequently,
we will use the asymptotic distribution with the delta method variance
as a working likelihood.\vadjust{\goodbreak}

\section{Combining data sources in the hierarchical Bayesian
space--time model}
\label{sec:models}

\subsection{The first stage}

Let $_5\widehat{q}_{0its}$ represent the estimate of U5MR from survey
$s$ in region $i$ and in period $t$.
A model-based approach to inference with survey data may be carried out
if the design variables upon which sampling were based, and associated
population totals, are available [\citet{gelman:07}]. Unfortunately,
these variables are not available for the Tanzania surveys. As an
alternative we summarize the data in area $i$ at time point $t$ from
survey $s$ via the asymptotic
distribution of the estimator of the pseudo-maximum likelihood
estimator (MLE):
\[
y_{its}=\log \biggl[ \frac{_5\widehat{q}_{0its}}{1- {_5\widehat
{q}_{0its}}} \biggr] .
\]
We define the area, period and survey summary
as $ \eta_{its}= \log [
_5q_{0its}/(1- {_5q_{0its}})
 ]$.
We take as working likelihood the asymptotic distribution
%
\begin{equation}
\label{eq:u5mlogit} y_{its} | \eta_{its}\sim\mathrm{N} (
\eta_{its}, \widehat {V}_{{\mathrm{DES}},its} ),
\end{equation}
which has been shown to perform well in the context of small area
estimation from complex surveys [\citet{mercer:etal:14}]. \citet
{dwyer:etal:2014} also used the pseudo-MLE, but did not incorporate
design effects and instead assumed a common variance across all
observations. However, Figure~13 from the supplementary material shows
that the variance of the five-year direct estimates can vary
significantly by survey and region.

\subsection{Second-stage smoothing models}

We wish to smooth over time period, region and surveys, but would like
as parsimonious a model as possible, to avoid overfitting.
At the second stage of our model we adopt a model similar to the ``Type
I'' inseparable space--time model of \citet{knorrheld:00}. However,
unlike \citet{knorrheld:00}, our data provides multiple observations for
each area $i$ and time point $t$ through the THMIS, five TDHS and two
HDSS, denoted as surveys~$s$. Thus, we consider models that allow the
option of survey-specific effects. The survey effects could be constant
over time and space, could vary with time, vary with space, or vary by
time and space.

The six candidate models we consider are given in Table~\ref{tab:random}, with the caption containing the random effects
specification. There are two temporal terms, with $\alpha_t$ being
independent and identically distributed random effects that pick up
short-term fluctuations with no structure, and $\gamma_t$ being given
an (intrinsic) random walk prior of order 1 or 2 (models type ``a'' or
``b''), to pick up local temporal smooth fluctuations, for $t=1,\ldots
,T=6$ time periods. Five-year time periods were chosen because
survey-specific regional sample sizes can be quite small. The UN IGME
has only recently moved to annual estimates at the national level
because the sample size of recent DHS has increased [\citet
{pedersenandliu:2012}]. We are combining recent and older DHS at a
regional level, and thus sample sizes are not sufficiently large to
produce reliable annual estimates.

\begin{table}
\caption{Random effects models for time period $t$, region $i$ and
survey $s$. In all models $\mu$ is the intercept and $\alpha_t\sim
_{\mathrm{i.i.d.}} \mathrm{N}(0,\sigma_\alpha^2)$, $\theta_i \sim_{\mathrm{i.i.d.}} \mathrm
{N}(0,\sigma_\theta^2)$, $\phi_i \sim\operatorname{ICAR}(\sigma_\phi^2)$,
$\delta_{it} \sim_{\mathrm{i.i.d.}} \mathrm{N}(0,\sigma_\delta^2)$. Specific models
contain random effects with distributions $\nu_s \sim_{\mathrm{i.i.d.}} \mathrm
{N}(0,\sigma_{\nu1}^2)$, $\nu_{is} \sim_{\mathrm{i.i.d.}} \mathrm{N}(0,\sigma
_{\nu
2}^2)$, $\nu_{ts} \sim_{\mathrm{i.i.d.}} \mathrm{N}(0,\sigma_{\nu3}^2)$, $\nu_{its}
\sim_{\mathrm{i.i.d.}} \mathrm{N}(0,\sigma_{\nu4}^2)$. In the ``a'' models
$\gamma_t
\sim\mathrm{RW1}(\sigma_\gamma^2)$ and in the ``b'' models $\gamma_t
\sim\mathrm{RW2}(\sigma_\gamma^2)$}\label{tab:random}
\begin{tabular*}{\textwidth}{@{\extracolsep{\fill}}lc@{}}
\hline
\multicolumn{1}{@{}l}{\textbf{Model}}&
\multicolumn{1}{c@{}}{\textbf{Linear predictor} $\bolds{\eta_{its}}$}\\
\hline
I&$\mu+\alpha_t+\gamma_t+\theta_i+ \phi_i +\delta_{it}$ \\
II&$\mu+\alpha_t+\gamma_t+\theta_i+ \phi_i +\delta_{it}+\nu_{s}$
\\
III&$\mu+\alpha_t+\gamma_t+\theta_i+ \phi_i +\delta_{it}+\nu
_{s}+\nu
_{is}$\\
IV&$\mu+\alpha_t+\gamma_t+\theta_i+ \phi_i +\delta_{it}+\nu
_{s}+\nu
_{ts}$\\
V&$\mu+\alpha_t+\gamma_t+\theta_i+ \phi_i +\delta_{it}+\nu
_{s}+\nu
_{ts}+\nu_{is}$\\
VI&$ \mu+\alpha_t+\gamma_t+\theta_i+ \phi_i +\delta_{it}+\nu
_{s}+\nu
_{ts}+\nu_{is}+\nu_{its}$\\
\hline
\end{tabular*}
\end{table}

There are also two spatial terms, corresponding to the convolution
model of \citet{besag:etal:91}. The independent random effects are
denoted $\theta_i$ and the intrinsic conditional autoregressive (ICAR)
terms are $\phi_i$ for $i=1,\ldots,I=21$ regions of Tanzania. The latter
perform local geographical smoothing. The space--time interaction terms
$\delta_{it}$ are taken to be independent, which corresponds to the
Type I interaction model of \citet{knorrheld:00}. Type II--IV interaction
models were considered, which include spatial and/or temporal structure
on the prior for $\delta_{it}$, but these models did not substantially
modify estimates, so Type I was selected for parsimony.

There are $S=8$ different surveys that are carried out over the various
time periods (since mothers are surveyed on their complete birth
history and so report on births from previous time periods), the five
TDHS and THMIS surveys cover all 21 regions over the different time
periods they were administered and the HDSS sites contribute data for
one region each in the last three time periods. The independent random
effects $\nu_s$ allow for these surveys to have a systematic
displacement from the true logit of U5MR. The interactions $\nu_{ts}$
and $\nu_{is}$ allow these displacements to vary with period and space,
respectively, while $\nu_{its}$ allow the complete interaction between
survey, period and area. Model I contains crossed random effects only,
since each area is represented in each of the time periods. Models
II--VI contain a combination of nested and crossed random effects.
The random walk and ICAR models are described in \citet{rue:knorrheld:05}.

\subsection{Hyperpriors}

For a generic set of independent random effects we specify priors on
the precision $\tau$ such that a 95\% prior interval for the residual
odds ratios lies in the interval $[0.5,2]$, which leads to
$\operatorname{Gamma}(a_{
{\mathrm{MARG}}},\break  b_{{\mathrm{MARG}}})$ priors for precisions
[\citet
{wakefield:09IJE}] with $a_{{\mathrm{MARG}}}=0.5$, $b_{{\mathrm
{MARG}}}=0.001488$. For the RW1, RW2 and ICAR models the precisions
have \textit{conditional} rather than \textit{marginal}
interpretations. Let $\bz$ represent a random effect from an improper
GMRF with ``mean'' $\bmzero$ and ``precision'' $\tau^\star\bQ$.
Following the supplementary material of \citet{fong:etal:10}, we gain
compatibility by calculating an approximate measure of the average
marginal ``variance'' of $\bz$ in the situation with $\tau^\star=1$;
call this average $c$. Then to put on the same scale, we take $a_{
{\mathrm{COND}}}=a_{{\mathrm{MARG}}}$ and $b_{{\mathrm
{COND}}}=b_{{\mathrm{MARG}}}/c$. In the above description, the words
mean, precision, and variance are written in italics to acknowledge
that, strictly speaking, these quantities do not exist since the
distribution is improper. However, one may calculate a generalized
inverse using the equation given at the end of Section~4.4 of \citet
{fong:etal:10}. This method is closely related to that later described
by \citet{sorbye:rue:14}. The supplementary material contain \texttt{R}
code for reproducing these prior specifications. For the Tanzania data
this leads to gamma priors for the RW1 of $\tau_\gamma\sim\operatorname
{Gamma}(0.5,0.00153)$, for the RW2 of $\tau_\gamma\sim\operatorname
{Gamma}(0.5,0.00286)$, and for the ICAR of $\tau_\phi\sim\operatorname
{Gamma}(0.5,0.00360)$.

\subsection{Computation}

Model fitting was carried out within the \texttt{R} computing environment.
Weighted logistic regressions were fit using the \texttt{svyglm()}
function from the \texttt{survey} package [\citet{lumley:04}] from which
the design-based variance was extracted (see supplementary material
for further details). The hierarchical Bayesian space--time models were
fitted using the Integrated Nested Laplace Approximation (INLA) [\citet
{rue:etal:09}] as implemented in the \texttt{INLA} package. INLA
provides a fast alternative to MCMC for approximating the marginal
posterior distributions of Markov random field (MRF) models. There is
now extensive evidence that the approximations are accurate for
space--time modeling; see for example \citet{fong:etal:10}, \citet
{held:etal:10} and \citet{schrodleheld:2011}.

\subsection{Model selection}\label{sec:select}

In Table~\ref{tab:random} we describe twelve plausible random effects
specifications (allowing for RW1 or RW2 models). A number of approaches
have been described for comparing models, including the conditional
predictive ordinate (CPO), the deviance information criteria (DIC) as
introduced by \citet{spiegelhalter:etal:02} and the normalizing
constants $p(\by|M)$ for the twelve models indexed by $M$.
Let $\by_{-its}$ represent the vector of data with the observation from
region $i$, time period $t$ and survey $s$ removed. The idea behind the
CPO is to predict the density ordinate of the left-out observation,
based on those that remain.
Specifically, the CPO for observation $i,t,s$ is defined as
\[
\mathrm{CPO}_{its} = p(y_{its}| \by_{-its}) = \int p(
y_{its} | \btheta) p(\btheta| \by_{-its} ) \,d\btheta =
E_{\theta|y_{-its}} \bigl[ p( y_{its} | \btheta) \bigr],
\]
where $\btheta$ represents the totality of parameters and in the U5MR
setting the distribution of $y_{its} | \btheta$ is N$(\eta
_{its},\widehat{V}_{\mathrm{DES},its})$. The CPOs can be used to
look at local fit or one can define an overall score for each model:
\[
\mathrm{LCPO} = \log\mathrm{(CPO)} = \sum_{i=1}^I
\sum_{t=1}^T \sum
_{s=1}^S\log\mathrm{CPO}_{its},
\]
and good models will have relatively high values of LCPO.
\citet{held:etal:10} discuss shortcuts for computation (i.e., avoidance
of fitting the model $I\times T\times S$ times) using INLA.

We also calculate another widely used model comparison measure, the
deviance information criteria, or DIC [\citet{spiegelhalter:etal:02}]. To
define the DIC with respect to a generic set of parameters $\btheta$,
first define an ``effective number of parameters'' as
\[
p_D = E_{\theta|y}\bigl\{ -2\log\bigl[p(\by|\btheta)\bigr]
\bigr\} + 2\log\bigl[ p(\by |\overline{\btheta})\bigr]= \overline{D} + D(
\overline{\btheta}),
\]
where $D$ is the deviance, $\overline{\btheta}=E[\btheta|\by]$ is the
posterior mean, $D(\overline{\btheta}) $ is the deviance evaluated at
the posterior mean and $\overline{D} = E[D|\by]$.
The DIC is given by
\[
\mathrm{DIC} = D(\overline{\btheta}) + 2 p_D= \overline{D} +
p_D,
\]
so that we have the sum of a measure of goodness of fit and model complexity.
We are wary of interpretation of DIC in our setting, since \citet
{plummer:08b} has shown that DIC is prone to inappropriately
under-penalize large models such as the ones we are fitting; see also
\citet{spiegelhalter:etal:14}.

\section{Applying methods to household surveys and HDSS sites in Tanzania}
\label{sec:modresults}

We fit models Ia--VIb (as summarized in Table~\ref{tab:random}) to the
Tanzania survey data and Table~\ref{tab:modelselect} provides the
summaries of various model comparison summaries. Model Vb is the
favored model according to both the DIC, LCPO, and log of the
normalizing constant criterion. Results for models Vb and VIb are very
similar, but we see from the effective number of parameters that even
though the number of 3-way interaction random effects is 573, there are
only 13 effective parameters due to the closeness of the interactions
to zero. Hence, from this point onward we shall report summaries with
respect to model Vb.
We begin by summarizing the posterior distribution, and then describe
regional trends.
%
\begin{table}
\caption{Model comparison: $p_D$ is the effective degrees of freedom,
as defined for the calculation of the deviance information criteria
(DIC), which also uses the deviance evaluated at the posterior mean,
$\overline{D}$; LCPO is defined as $\sum_{its}\operatorname
{log}(\mathrm{CPO}_{its})$. In the ``a'' models $\gamma_t \sim\mathrm
{RW1}(\sigma
_\gamma^2)$ and in the ``b'' models $\gamma_t \sim\mathrm{RW2}(\sigma
_\gamma^2)$}\label{tab:modelselect}
\begin{tabular*}{\textwidth}{@{\extracolsep{\fill}}lccd{3.1}ccc@{}}
\hline
\textbf{Model} & \textbf{No pars}&\multicolumn{1}{c}{$\bolds{\log p(\by)}$}&
\multicolumn{1}{c}{$\bolds{p_D}$}&
\multicolumn{1}{c}{$\bolds{\overline{D}}$} & \textbf{DIC}&\textbf{LCPO} \\
\hline
Ia & 181& $-$297.3 & 74.5 & 409.3 & 483.8 & $-$294.5 \\
IIa& 189& $-$291.0 & 80.1 & 384.2 & 464.3 & $-$287.3 \\
IIIa& 313& $-$244.1 & 118.9 & 221.8 & 340.7 & $-$193.5 \\
IVa& 223& $-$288.6 & 88.6 & 367.5 & 456.2 & $-$283.4 \\
Va& 347& $-$241.2 & 121.8 & 210.1 & 332.0 & $-$183.1 \\
VIa& 920 & $-$241.4 & 134.5 & 199.4 & 334.0 & $-$183.9 \\ [3pt]
Ib & 181& $-$293.3 & 74.2 & 409.1 & 483.3 & $-$293.7 \\
IIb& 189& $-$287.0 & 79.8 & 383.9 & 463.7 & $-$286.4 \\
IIIb& 313& $-$239.9 & 118.6 & 221.7 & 340.3 & $-$192.9 \\
IVb& 223& $-$284.5 & 88.2 & 367.4 & 455.6 & $-$282.5 \\
Vb& 347&\textbf{$\bolds{-}$236.9} & 121.6 & 209.9 & \textbf{331.5} &
\textbf{$\bolds{-}$183.1} \\
VIb& 920& $-$237.6 & 133.3 & 200.2 & 333.4 & $-$183.4 \\
\hline
\end{tabular*}
\end{table}

\subsection{Summarizing the posterior distribution}

Table~\ref{tab:varcomps} provides numerical summaries and the
proportion of total variation explained by each random effect. The
total variance is
\[
\sigma_\alpha^2+s_\gamma^2+
\sigma_\theta^2+s_\phi^2+\sigma
_\delta ^2+\sigma_{\nu_s}^2+
\sigma_{\nu_{si}}^2+\sigma_{\nu_{st}}^2,
\]
where $s_\gamma^2$ and $ s_\phi^2$ are empirical estimates of the
marginal variances in the RW2 and ICAR models.
The structured temporal and unstructured spatial random effects explain
77\% of the total variation. Hence, there is strong temporal structure
and large spatial heterogeneity, which we shall discuss subsequently.
The third largest contribution to the variation is 11\% from the
survey--space interaction. Different survey teams are sent to different
regions, which explains to some extent this relatively large contribution.

\begin{table}[b]
\caption{Summaries of variance components. The proportion of variation
is calculated as the contribution the relevant set of random effects
makes to the total variation. In the case of the RW2 and ICAR models,
the relevant contribution is evaluated empirically, since the variance
parameter is conditional rather than marginal}\label{tab:varcomps}
\begin{tabular*}{\textwidth}{@{\extracolsep{\fill}}lccd{2.1}@{}}
\hline
\textbf{Variance}&\textbf{Interpretation}&
\textbf{Median (95\% interval)}&\multicolumn{1}{c@{}}{\textbf{Percentage variation}}\\
\hline
$\sigma_\alpha^2$&Indept time&0.002 (0.001, 0.012) & 1.3\\[2pt]
$\sigma_\gamma^2$&RW2 time&0.009 (0.002, 0.054) & 46.0\\[2pt]
$\sigma_\theta^2$&Indept space&0.068 (0.033, 0.133) & 31.3\\[2pt]
$\sigma_\phi^2$&ICAR space&0.017 (0.002, 0.378) & 4.9\\[2pt]
$\sigma_\delta^2$&Indept space--time interaction&0.005 (0.001, 0.013) &
2.3\\[2pt]
$\sigma_{\nu_s}^2$&Indept survey&0.002 (0.001, 0.013) & 1.4\\[2pt]
$\sigma_{\nu_{st}}^2$&Indept survey--time interaction&0.004 (0.001,
0.011) & 2.0\\[2pt]
$\sigma_{\nu_{si}}^2$&Indept survey--space interaction&0.024 (0.015,
0.038) & 10.9\\
\hline
\end{tabular*}
\end{table}

\subsection{Model validation}\label{sec:validate}

To validate the model, we removed all of the observations in area $i$
for time point $t$ and then generated 95\% intervals around the
posterior mean $_5\tilde{q}_{0,it}$
using the variance of the observed response, defined as $\tilde
{S}^2_{its}=\tilde{\sigma}^2_{it}+\widehat{V}_{\mathrm{{DES,its}}}$,
where $\tilde{\sigma}^2_{it}$ is the variance of the posterior
distribution of $\operatorname{logit} (_5\tilde{q}_{0,it} )$ and $\widehat
{V}_{\mathrm{{DES,its}}}$ is the design-based variance described in
Section~\ref{sec:dtsa}. This was completed for the 21 regions and 6
time points (figures shown in the supplementary material). Intervals
contained the design-based estimates 92.5\% of the time overall.
Time/area-specific coverages range from 89.9--96.9\% and the coverage
for the final time point is 93.2\%.

\begin{figure}

\includegraphics{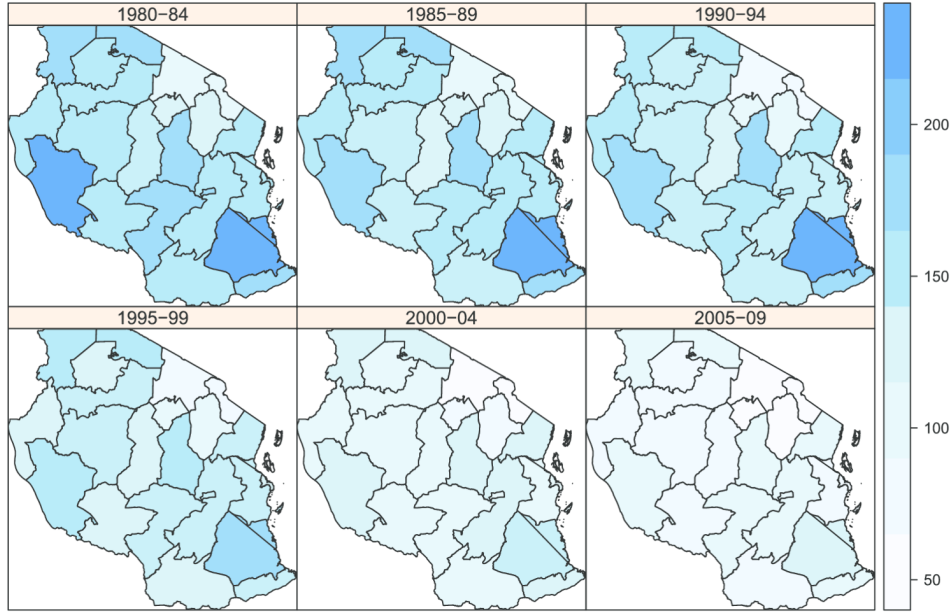}

\caption{The solid line represents five-year model-smoothed estimates of
$_5q_0$ in Pwani region, TZA, with 95\% confidence intervals as vertical lines. The dashed lines
display the five-year direct estimates from the region by household survey and surveillance
site, with 95\% confidence intervals as vertical lines.}\vspace*{-12pt}
\label{fig:smoothed_maps}
\end{figure}

\begin{figure}

\includegraphics{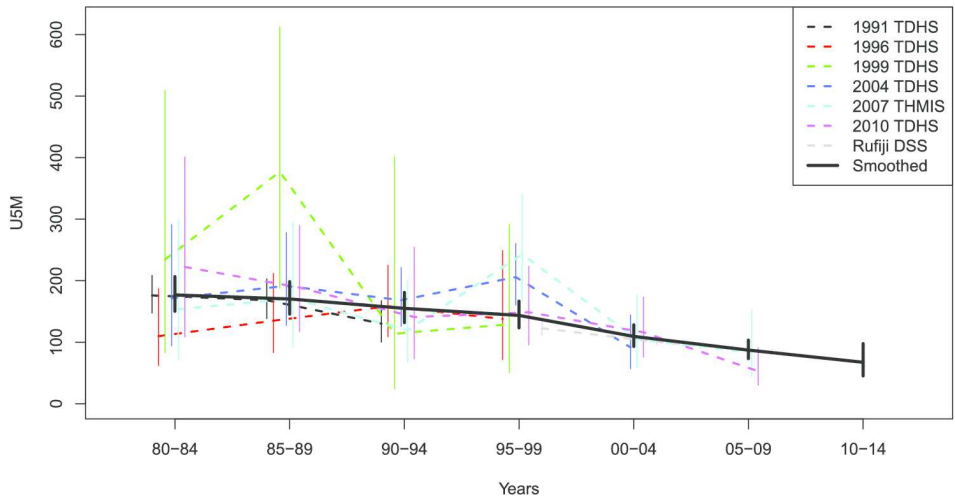}

\caption{Regional five-year direct and model-based smoothed of $_5q_0$
in Pwani, TZA, with 95\% confidence intervals.}
\label{fig:pwani}
\end{figure}

\subsection{Regional estimates and projections}

\begin{figure}

\includegraphics{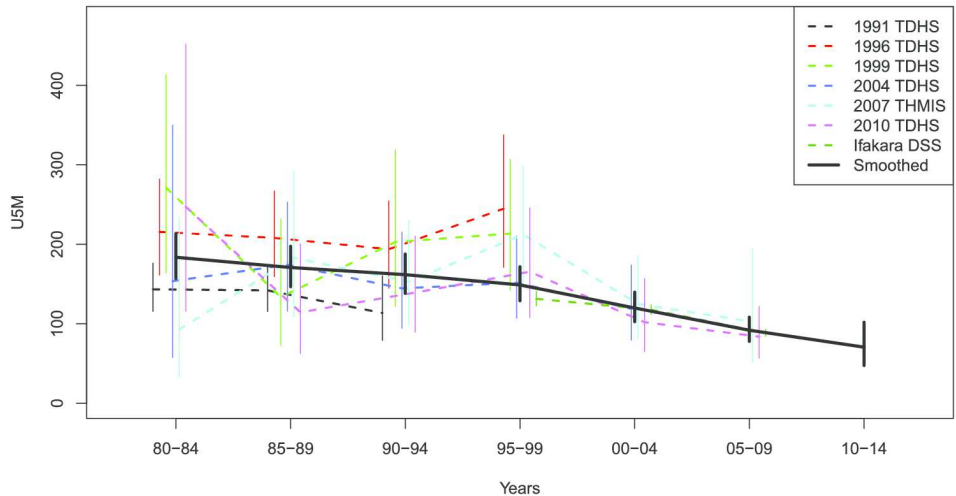}

\caption{The solid line represents five-year model-smoothed estimates of $_5q_0$
in Morogoro region, TZA, with 95\% confidence intervals as vertical lines. The dashed lines
display the five-year direct estimates from the region by household survey and surveillance
site, with 95\% confidence intervals as vertical lines.}
\label{fig:morogoro}
\end{figure}

For region $i$ and 5-year period $t$, estimates, projections and
credible intervals of U5MR are taken from posterior draws of
\[
_5q_{0,it}=\operatorname{expit} ( \mu+\alpha_t+
\gamma_t+\theta _i+\phi_i +
\delta_{it} ).
\]
Figure~\ref{fig:smoothed_maps} shows maps of the posterior median
estimates of child mortality (per 1000 births) by region for the six
observed 5-year time periods. Child mortality has decreased markedly
over the 30-year period considered, but overall more than 5\% of
infants still die before they turn 5, and there are strong regional
differences. Figures \ref{fig:pwani} and \ref{fig:morogoro} display the
observed direct estimates and smoothed results for the Morgoro and
Pwani regions, respectively. Additionally, each plot shows the
projected U5MR for the 2010--2014 time period. The direct estimates
have a great deal of variability between surveys, especially for the
first four time points, and design-based intervals are very wide.
Smoothed rates and projections for all 21 regions are located in the
supplementary material.

\section{Discussion}\label{sec:discuss}

We have described a general method for
spatiotemporal smoothing of a health outcome, with the data arising
from complex surveys and surveillance.\vadjust{\goodbreak} The method was illustrated with
child mortality in regions of Tanzania over 1980--2009 using data from
household surveys and surveillance sites. A great advantage of the
model is that there is a fast implementation within the R computing
environment using the existing \texttt{survey} and \texttt{INLA}
packages. The supplementary material contain example code. As an
example, fitting the most complex model for the Tanzania data took
just 18.7 seconds on a Macbook Pro.\setcounter{footnote}{2}\footnote{Processor: 2.9 GHz Intel
Core i7; memory: 8 GB 1600 MHz DDR3.}

In our hierarchical modeling approach, we explicitly acknowledge the
weights by taking as (pseudo-)likelihood the (design-based) sampling
distribution of the estimator. In the supplementary material we
illustrate the effect of the weights on both the estimates and the
standard errors. Another use of our model is
for prediction, with the RW2 terms drawn from the relevant conditional
distribution.

Our model contains a relatively complex combination of nested and
crossed random effects and we described a particular approach to
hyperprior selection. As with any such suggestion, it is beneficial to
examine prior sensitivity, and the supplementary material contain
details of a sensitivity study that we performed for the Tanzania data.

An integral part of our
method involves calculating and pooling estimates of child mortality
from household surveys and demographic surveillance sites and allowing
both to inform our overall estimates by region and for the country as
a whole. A byproduct of this procedure is an ability to carefully
compare the DHS-based and demographic surveillance-based estimates of
child mortality in the regions that include HDSS sites. As Figures \ref{fig:pwani} and \ref{fig:morogoro} make clear, the central
estimates from the two different data
collection schemes are very similar. This adds more weight to similar
findings by others
[\citet{byass2007dss,fottrell2009distribution,hammer2006risk}] and
reduces concerns about the Hawthorne effect preventing measures of
child mortality from HDSS sites from being more widely relevant,
that is, similar to surrounding populations.

Although we have demonstrated our method with a single country and
outcome, it is
sufficiently general to be applied to produce spatiotemporal estimates
of a variety of indicators. Because this approach provides
consistent, precise estimates across both time and space utilizing
data from a variety of sources, including complex sample surveys,
accounting for study designs, it should be considered as an
approach for producing subnational estimates of child
mortality and other key health, demographic and development
indicators. However, countries with a substantial HIV/AIDs burden may
suffer from underreporting biases.
The UN IGME preprocesses data in a number of countries, including
Tanzania, to take account of underreporting biases because of HIV/AIDS.
We base our analysis on direct subnational estimates of U5MR, and so do
not adjust for this bias, but our smoothed results do not differ
substantially from the UN results at the national level and so we
believe that any bias from this source will be small.

The world's rapidly growing appetite for timely, subnational estimates
of key development indicators will continue to motivate innovative new
developments in both data collection and analysis. In addition to
providing a means to improve indicator estimates using different
sources of data, our results also hint at the possibility of
eventually creating integrated data collection and analysis schemes
that build on existing infrastructure to yield some of the
functionality of full-coverage CRVS. \citet{hyak} and
\citet{ye2012health} begin to discuss ideas in this vein, for example,
how one
might utilize both sample surveys and demographic surveillance to
continuously provide indicators equivalent to what is normally
produced by vital registration. The method and results we present in
this paper encourage future development of those ideas.

\section*{Acknowledgments}
We would like to acknowledge the hard work and commitment of the HDSS
field teams and the data team, and thank the residents in the
surveillance areas for offering their time for interviews and sharing
their valuable personal information over the years.


\begin{supplement}[id=suppA]
\stitle{Supplement to ``Space--time smoothing models for complex
survey data: Small area estimation for child mortality''}
\slink[doi]{10.1214/15-AOAS872SUPP} 
\sdatatype{.pdf}
\sfilename{aoas872\_supp.pdf}
\sdescription{The organization of the supplementary material is as follows. In Section 1 we provide the
details of the discrete survival model. In Section 2 we provide the derivation of the standard
error for U5M. Section 3 describes a simulation study aimed to test the coverage performance
of the derived standard error against the jackknife standard error used by DHS. In Section
4 we describe the hyperprior specifications for the Bayesian hierarchical model. Section 5
provides a summary of the posterior distribution of the random effects. In Section 6 we
provide a comparison of weighted and unweighted direct estimates of U5M. In Section 7
we have included some exploratory analysis looking at the rates and magnitude of regional
decreases in U5M and how they relate to the fourth millennium development goal of two
thirds reduction in child mortality by 2015. The results of our model validation are presented
in Section 8. Lastly, Section 9 includes example R code for the analyses.}
\end{supplement}



%




\printaddresses
\end{document}